\documentclass[final,1p,times]{elsarticle}
\biboptions{sort&compress}
\usepackage{amsmath,amssymb}
\usepackage{amsthm}
\usepackage{bbm}
\usepackage{bm}
\usepackage{psfrag}
\usepackage{graphicx}

\newcommand\RR{{\mathbbm{R}}}

\newcommand\dd{{\mathrm{d}}}

\newcommand\ii{{\mathrm{i}}}

\newcommand{\mvec}[1]{\bm{#1}}
\DeclareMathOperator{\sgn}{sgn}
\DeclareMathOperator{\Tr}{Tr}
\DeclareMathOperator{\Real}{Re}

\newtheorem*{hdc}{Criterion}
\newtheorem{theorem}{Theorem}

\begin{document} 

\begin{frontmatter}

\title{Stationary point analysis of the one-dimensional lattice Landau gauge fixing functional, aka random phase $XY$ Hamiltonian} 

\author[addDM]{Dhagash Mehta}
\ead{dbmehta@syr.edu}

\author[addMK]{Michael Kastner} 
\ead{kastner@sun.ac.za} 


\address[addDM]{Department of Physics, Syracuse University, Syracuse NY 13244, USA}

\address[addMK]{National Institute for Theoretical Physics (NITheP), Stellenbosch 7600, South Africa, and\\
Institute of Theoretical Physics,  University of Stellenbosch, Stellenbosch 7600, South Africa}

 
\begin{abstract}
We study the stationary points of what is known as the lattice Landau gauge fixing functional in one-dimensional compact U(1) lattice gauge theory, or as the Hamiltonian of the one-dimen\-sion\-al random phase $XY$ model in statistical physics. An analytic solution of all stationary points is derived for lattices with an odd number of lattice sites and periodic boundary conditions. In the context of lattice gauge theory, these stationary points and their indices are used to compute the gauge fixing partition function, making reference in particular to the Neuberger problem. Interpreted as stationary points of the one-dimensional $XY$ Hamiltonian, the solutions and their Hessian determinants allow us to evaluate a criterion which makes predictions on the existence of phase transitions and the corresponding critical energies in the thermodynamic limit.
\end{abstract}

\end{frontmatter}

\section{Introduction}
\label{sec:intro}

Stationary points, i.\,e.\ points of vanishing gradient of some function, play a prominent role in many methods of theoretical physics. Generally speaking, any variational principle uses stationarity to single out a certain point in a (function) space. One field of physics where stationary points feature prominently is gauge fixing via a gauge fixing functional. Recently, a variety of methods summarized under the name of ``energy landscape methods'' \cite{Wales} have attracted quite some attention, allowing for applications to many-body systems as diverse as metallic clusters, biomolecules and their folding transitions, or glass formers undergoing a glass transition. In all these examples, the stationary points of some (usually high-dimensional) energy (or potential energy, or free energy) function are studied. In most cases such stationary point approaches are used in numerical studies. Exact, analytic studies of the stationary points are rarely feasible but, as is often the case, can be of great value, as they may prove useful for verifying ideas or testing approximation schemes, and they may be instructive towards the understanding of more complicated situations.

In some instances, it may happen that not only the same method of analyzing stationary points is used for applications in different fields of physics, but even the same function is analyzed, albeit in a very different context. In this article, we want to report such a case, which is an analysis of the stationary points of the function
\begin{equation}\label{eq:F_phi}
F_\phi (\theta)=\sum_{k=1}^N [1-\cos(\phi_k +\theta_{k+1}-\theta_k)].
\end{equation}
The function $F_\phi$ is defined on an $N$-torus, parametrized by the angular variables $\theta=(\theta_1,\dots,\theta_N)$ with $\theta_k\in(-\pi,\pi]$. The random parameters $\phi_k\in(-\pi,\pi]$ introduce a quenched disorder, and we will be looking for stationary points of $F_\phi$ with respect to the $\theta$-variables,
\begin{equation}\label{eq:gaugefixing}
 0 = \frac{\partial F_\phi (\theta)}{\partial \theta_k} = \sin(\phi_{k-1} +\theta_k-\theta_{k-1}) - \sin(\phi_k +\theta_{k+1}-\theta_k),\qquad k=1,\dots,N,
\end{equation}
valid for arbitrary fixed values of $\phi=(\phi_1,\dots,\phi_N)$. A central result of this paper is an exact, analytic expression for the stationary points of this function for periodic boundary conditions. Depending on the problem under investigation, various properties of the stationary points are used in the analysis and related to physical quantities. This may be the number of stationary points, their respective energy value, the index of a stationary point or its Hessian determinant.

In physics, the function $F_\phi$ given in \eqref{eq:F_phi} appears in at least two different contexts: First, it is the lattice Landau gauge functional for a compact U(1) lattice gauge theory in one spatial dimension. Each of the stationary points, i.\,e.\ each solution of the set of equations \eqref{eq:gaugefixing}, corresponds to a gauge fixing, and a number of interesting phenomena like the Gribov ambiguity and the Neuberger problem are related to the stationary points and their properties. Second, in statistical physics the function $F_\phi$ is known as the Hamiltonian of the classical random-phase $XY$ model in one spatial dimension. This model consists of $N$ classical planar spin variables on a one-dimensional lattice where each spin is coupled to its two nearest neighbors on the lattice. In recent years, there has been significant progress towards establishing a relation between stationary points of classical many-body Hamiltonians and the occurrence or absence of phase transitions, and it is in this spirit that we will analyze the stationary points of the $XY$ Hamiltonian in this article. Since we assume that the average reader will be familiar with at most one of these fields of physics, but not both, we will provide elementary introductions in later sections.

The main concern of the present article is twofold: First, the exact, analytic solution of the stationary points of $F_\phi$ provides a testing ground for ideas, methods, and approximations that are used for higher-dimensional or otherwise more complicated models where exact results are not available. Second, juxtaposing an interpretation of the stationary points as gauge fixing of a compact U(1) lattice gauge theory with an interpretation as special points of a classical spin Hamiltonian beautifully illustrates the similarity as well as the differences of stationary point approaches in two very different fields of physics.

The article is structured as follows: In section \ref{sec:stationarypoints}, an exact, analytical expression is derived for the stationary points of $F_\phi$ with periodic boundary conditions. A short introduction to gauge fixing in lattice gauge theory, the Gribov ambiguity, and the Neuberger problem is given in section \ref{sec:SLLG}. In section \ref{sec:gaugefixing}, $F_\phi$ is shown to be the lattice Landau gauge functional of the one-dimensional, compact U(1) lattice gauge theory, and the results from section \ref{sec:stationarypoints} are used to discuss the Neuberger problem for this case. Possible strategies to avoid the Neuberger problem and the Gribov ambiguity are discussed in section \ref{sec:modifiedLLG}. In section \ref{sec:XY} we turn to the interpretation of $F_\phi$ as the Hamiltonian of a classical random-phase $XY$ spin system. In section \ref{sec:phasetransitions}, some recent results on stationary points and their relation to the occurrence of phase transitions are sketched. In section \ref{sec:PTXY} the random phase $XY$ model is introduced and some thermodynamic properties of the one-dimensional $XY$ model are sketched. In section \ref{sec:statpoint1dXY}, the stationary points of the Hamiltonian of this model are put in relation to its thermodynamic behavior. A summary and an outlook on open questions and possible future work is given in section \ref{sec:summary}.


\section{Calculating stationary points}
\label{sec:stationarypoints}

The stationary points of the function $F_\phi$ as defined in equation \eqref{eq:F_phi} depend crucially on the conditions imposed on the variables $\theta_1$ and $\theta_N$ at the boundaries. In this section, the stationary points of $F_\phi$ are calculated for periodic boundary conditions. The stationary points of $F_\phi$ with anti-periodic boundary conditions have been computed previously in \cite{vonSmekal:2007ns}. Remarkably, the solutions found in the two cases differ substantially. For anti-periodic boundary conditions, two fairly simple classes of solutions of the set of equations \eqref{eq:gaugefixing} were shown in \cite{vonSmekal:2007ns} to exhaustively comprehend the $2^N$ stationary points of $F_\phi$, all of which are isolated. For periodic boundary conditions, the global rotational symmetry, $\theta_k\to\theta_k+\alpha$ for any real $\alpha$ and $k=1,\dots,N$, trivially accounts for continuous curves of solutions. But even after fixing this global invariance, the solutions for the stationary points, as will be shown below, are significantly more variegated than in the case of anti-periodic boundary conditions.

To compute the stationary points of $F_\phi$ with periodic boundary conditions, it is convenient to transform the set of equations \eqref{eq:gaugefixing} to the new angular variables
\begin{equation}\label{eq:defsk}
(-\pi,\pi]\ni s_k:=\phi_k+\theta_{k+1}-\theta_k \mod 2\pi,
\end{equation}
yielding
\begin{equation}\label{eq:s_gaugefixing}
0=\sin s_{k+1} - \sin s_k,\qquad k=1,\dots,N-1,
\end{equation}
together with
\begin{equation}\label{eq:s_gaugefixing_boundary}
0=\sin s_1 - \sin s_{N}.
\end{equation}
Equivalently, we can write these equations in matrix form,
\begin{equation}\label{eq:MS}
M\mvec{S}=\mvec{0},
\end{equation}
with column vector $\mvec{S}=(\sin s_1,\dots,\sin s_N)$ and zero vector $\mvec{0}$, where $M$ is a circulant $(N\times N)$-matrix of the form
\begin{equation}
M=\begin{pmatrix}
-1&1&0&0&\cdots&0\\
0&-1&1&0&\cdots&0\\
0&0&-1&1&\ddots&\vdots\\
\vdots&\vdots&\ddots&\ddots&\ddots&0\\
0&0&\cdots&0&-1&1\\
1&0&\cdots&0&0&-1
\end{pmatrix}.
\end{equation}
Like for any circulant matrix, the eigenvalues of $M$ can be readily obtained by a discrete Fourier transform of an arbitrary row or column of the matrix \cite{Gray06}. One of the eigenvalues is found to be zero, corresponding to the invariance of the set of equations \eqref{eq:s_gaugefixing} and \eqref{eq:s_gaugefixing_boundary} under the abovementioned global phase shift. As a consequence of this invariance, the solutions of \eqref{eq:MS} are not isolated, but occur in one-parameter families. Moreover, only $N-1$ out of the $N$ equations in \eqref{eq:s_gaugefixing} and \eqref{eq:s_gaugefixing_boundary} are linearly independent. To get rid of this trivial invariance, we can set, without loss of generality, $\theta_{N}=0$, and eliminate equation \eqref{eq:s_gaugefixing_boundary}. Since the inverse function of the sine is multi-valued, we obtain multiple solutions for the $s_k$ in \eqref{eq:s_gaugefixing},
\begin{equation}\label{eq:s_solutions}
s_k+2\pi l_k = s_N \qquad\text{or}\qquad s_k+2\pi l_k = \pi-s_N,
\end{equation}
with $l_k$ being integer valued. Equivalently, we can write
\begin{equation}\label{eq:s_solutions2}
s_k+2\pi l_k = (-1)^{q_k}s_N+q_k\pi,\qquad q_k\in\{0,1\}.
\end{equation}
Summing the $N-1$ equations in \eqref{eq:s_solutions2}, we obtain
\begin{equation}\label{eq:sum_s_N}
\sum_{k=1}^{N-1}(s_k+2\pi l_k-\pi q_k) = s_N\sum_{k=1}^{N-1}(-1)^{q_k}.
\end{equation}
Making use of the definitions $s_k=\phi_k+\theta_{k+1}-\theta_k$ for $k=1,\dots,N-1$ and $s_N=\phi_N+\theta_{1}-\theta_N$, we can write
\begin{equation}
\sum_{k=1}^{N-1}s_k=\theta_N-\theta_{1}-\phi_N+\sum_{k=1}^N\phi_k=-s_N+\Phi,
\end{equation}
where we have defined $\Phi:=\sum_{k=1}^{N}\phi_{k}$. Inserting this expression into \eqref{eq:sum_s_N} and solving for $s_N$, one obtains
\begin{equation}
s_N=\frac{\Phi-\pi\sum_{k=1}^{N-1}q_k}{1+\sum_{k=1}^{N-1}(-1)^{q_k}}+\frac{2\pi\sum_{k=1}^{N-1}l_k}{1+\sum_{k=1}^{N-1}(-1)^{q_k}},
\end{equation}
valid for $1+\sum_{k=1}^{N-1}(-1)^{q_k}\neq0$. In case the denominator vanishes, we do not obtain a well defined solution for $s_N$. 
Although we have little doubt that this issue could be resolved, we will simply avoid the problem of vanishing denominators by restricting $N$ to odd integers. From the second term on the right hand side we can infer that the value of $s_N$ does not really depend on the specific sequence of $l_1,\dots,l_{N-1}$, but only on their sum. Without loss of generality, we can therefore write
\begin{equation}\label{eq:s_final}
s_N=\frac{\Phi+2\pi l-\pi\sum_{k=1}^{N-1}q_k}{1+\sum_{k=1}^{N-1}(-1)^{q_k}},
\end{equation}
where $l$ is an integer valued parameter. In fact, since we are interested in solutions of $s_N$ only modulo multiples of $2\pi$, it is sufficient to consider
\begin{equation}\label{eq:l_range}
l\in\biggl\{1,\dots,\Bigl|1+\sum_{k=1}^{N-1}(-1)^{q_k}\Bigr|\biggr\}.
\end{equation}
Summarizing, for $N$ odd, we have obtained an expression for $s_N$, parametrized by the sequence $(q_1,\dots,q_{N-1})$ with $q_k\in\{0,1\}$ and by the parameter $l$. To comply with the convention fixed in \eqref{eq:defsk}, $s_N$ should then be projected onto the interval $(-\pi,\pi]$. The sequence of variables $(s_1,\dots,s_{N-1})$ is readily obtained by making use of equation \eqref{eq:s_solutions2}; here $l_k=0$ for all $k=1,\dots,N-1$ can be used, since we are only interested in solutions for the $s_k$ modulo multiples of $2\pi$. Making use of $\theta_N=0$, it is then straightforward to translate the result to the original variables $\theta_k$ by inverting the definitions \eqref{eq:defsk}.

Among these solutions, there are two particularly simple classes of stationary points: (a) any sequence $(s_1,\dots,s_N)$ where $s_k\in\{\Phi,\Phi+\pi\}$ for all $k$ is a stationary point of $F_\phi$; and (b) any sequence of identical $s_k\equiv (\Phi+2\pi l)/N$ with $l=1,\dots,N$.

The number $\#(\theta^\text{s})$ of stationary points $\theta^\text{s}$ that exist for a given number $N$ of lattice sites is obtained straightforwardly: Consider all sequences $(q_1,\dots,q_{N-1})$ consisting of a certain fixed number $j$ of 1-entries, and $N-j-1$ zero entries. The number of such permutations is $\binom{N-1}{j}$ and, according to \eqref{eq:l_range}, the parameter $l$ can take on values ranging from zero to $|N-2j|$. Summing over all possible values of $j$, we obtain the overall number of stationary points,
\begin{equation}\label{eq:number}
\#(\theta^\text{s})=\sum_{j=0}^{N-1}|N-2j|\,\binom{N-1}{j}=\frac{N!}{[(N-1)/2)!]^2},
\end{equation}
where the final expression follows after some tedious but rather straightforward algebra. Not unexpectedly \cite{DoyeWales02}, asymptotically for large system sizes $N$, applying Stirling's formula to \eqref{eq:number} shows that the number of stationary points increases exponentially with $N$. The number of minima of $F_\phi$ can be obtained by reverting to the following result.
\begin{theorem}\label{theorem}
Consider the function $F_\phi$ as defined in \eqref{eq:F_phi} and its stationary points as derived above, each of which is parametrized by a sequence $(q_1,\dots,q_N)$ with $q_k\in\{0,1\}$ and by $l\in\{1,\dots,1+\sum_{k=1}^{N-1}(-1)^{q_k}\}$. A stationary point of $F_\phi$ is a minimum if and only if
\begin{enumerate}
\item[(a)] $q_k=0$ for all $k=1,\dots,N-1$, and
\item[(b)] $\cos s_N>0$, with $s_N$ as defined in \eqref{eq:s_final}.
\end{enumerate} 
\end{theorem}
The proof of this result is similar to a related statement shown in the Appendix of \cite{CaPeCo:03}, and we refer the reader to this article for more information on the strategy of proof. Note that, via the $\Phi$-dependence of $s_N$, the number of minima of $F_\phi$ depends in general on the quenched disorder $\phi=(\phi_1,\dots,\phi_N)$. Interestingly, one can read off from the above theorem that the number of minima is bounded from above by $N$. The number of minima increases therefore at most linearly with $N$, but not exponentially, in contrast to the general expectation based on an assumption of independent subsystems \cite{DoyeWales02}. 

Evaluating the function \eqref{eq:F_phi} at a stationary point, we obtain
\begin{equation}\label{eq:F_at_SP}
F_\phi(\theta^\text{s}) = N-\cos s_N\sum_{k=1}^{N}(-1)^{q_k}.
\end{equation}
The Hessian $\mathcal{H}=(\mathcal{H}_{ij})$ of $F_\phi$ is a tridiagonal matrix with elements
\begin{equation}\label{eq:Hessian}
\mathcal{H}_{ij}=\frac{\partial^2 F_\phi(\theta)}{\partial \theta_i \partial \theta_j} = \delta_{i,j}(\cos s_{i-1}+\cos s_i)-\delta_{i-1,j}\cos s_{i-1}-\delta_{i+1,j}\cos s_i,\qquad i,j=1,\dots,N-1.
\end{equation}
Note that $\mathcal{H}$ is an $(N-1)\times(N-1)$-matrix, accounting for the fact that we have removed one linearly dependent equation out of the $N$ equations \eqref{eq:s_gaugefixing} and \eqref{eq:s_gaugefixing_boundary}. The Hessian determinant can be computed explicitly, yielding 
\begin{equation}\label{eq:Hessian_determinant}
 \det \mathcal{H} =\sum_{j=1}^{N} \prod_{\substack{k=1\\k\neq j}}^{N}\cos s_k.
\end{equation}
Evaluating $\mathcal{H}$ at the stationary points $\theta^\text{s}$ of $F_\phi$, we find
\begin{equation}\label{eq:HessianDetAtSP}
\det \mathcal{H}(\theta^\text{s})=(\cos s_N)^{N-1}\Biggl(1+\sum_{k=1}^{N-1}(-1)^{q_k}\Biggr)\Biggl(\prod_{j=1}^{N-1}(-1)^{q_j}\Biggr).
\end{equation}


\section{Lattice Landau gauge fixing}
\label{sec:SLLG}

In this section, a brief introduction to gauge fixing on the lattice is given. In section \ref{sec:gaugefixing}, the function $F_\phi$ defined in \eqref{eq:F_phi} will be identified as the Landau gauge fixing functional of a one-dimensional compact U(1) lattice gauge theory, and this connection allows us to use the stationary points of $F_\phi$ computed in section \ref{sec:stationarypoints} for studying the gauge fixing partition function.

A gauge field theory can be successfully studied nonperturbatively by discretizing the Euclidean space-time and putting the gauge and matter fields of the theory on a $(d+1)$-dimensional space-time grid \cite{Rothe:2005nw}. A lattice grid consists of lattice sites, as well as links connecting any two adjacent sites. The gauge fields are usually defined through link variables $U_{i,\mu}\in G$ where $i$ denotes the site index, $\mu$ is a directional index, and $G$ is the gauge group of the theory. By $U=\{U_{i,\mu}\}$ we define the set of link variables for all $i$ and $\mu$. These link variables are related to the gauge fields of the continuum theory by the standard Wilson discretization \cite{Rothe:2005nw}.

Gauge fixing is a procedure to get rid of the redundant degrees of freedom of the field variables in gauge theories. Technically, gauge fixing is implemented by requiring the gauge fields to satisfy a certain constraint, like for example one of the well-known gauge conditions in classical electrodynamics. In a continuum gauge theory, gauge fixing is indispensable in order to assure convergence of certain integrals when computing expectation values of observables. In lattice gauge theory, the spatial discretization renders the theory manifestly gauge invariant: when computing expectation values of an observable $O$ with respect to the (gauge-invariant) Wilson action $S$,
\begin{equation}\label{eq:exp_value_of_O_lattice}
 \langle O \rangle = \frac{\int \prod_{i,\mu}d U_{i,\mu} \exp(-S[U])O[U]}{\int \prod_{i,\mu}d U_{i,\mu} \exp(-S[U])}.
\end{equation}
the occurring integrals are well-defined, and fixing a gauge is therefore not required. Still, fixing a gauge on the lattice can be beneficial, for example when comparing with gauge dependent results from the corresponding continuum theory. For example, in continuum gauge theories, gauge-dependent quantities are at the basis of certain confinement scenarios such as Kugo-Ojima and Gribov-Zwanziger~\cite{Kugo:1979gm, Gribov:1977wm}, or Dyson-Schwinger equation studies for the quantum chromodynamics hadron phenomenology \cite{Alkofer:2000wg}.

In its modern formulation, a gauge fixing condition is usually specified in terms of the minima or, more generally, the stationary points of some functional \cite{vonSmekal:2007ns, vonSmekal:2008es, vonSmekal:2008ws, Mehta:2009}. On the lattice, a popular choice is the lattice Landau gauge functional \cite{Davies:1987vs},
\begin{equation}\label{eq:general_l_g_functional}
 F_{U}(g) = \sum_{i,\mu}\left(1-
\Real \Tr g_{i}^{\dagger}U_{i,\mu}g_{i+\hat{\mu}} \right),
\end{equation}
where $g=(g_1,g_2,\dots)$ with $g_i\in G$ are gauge transformations, and the summation in \eqref{eq:general_l_g_functional} runs over all sites $i$ and all link directions $\mu$ on the lattice. The gauge-fixing conditions are then defined as the vanishing first derivatives of $F_{U}$ with respect to the gauge transformations $g$,
\begin{equation}\label{eq:gaugefixing_gen}
f(g)\equiv(f_1(g),f_2(g),\dots)=0 \qquad\text{with}\qquad f_{i}(g) := \frac{\partial F_{U}(g)}{\partial g_{i}}.
\end{equation}
The functional \eqref{eq:general_l_g_functional} is chosen such that the gauge-fixing equations \eqref{eq:gaugefixing_gen} reproduce the usual continuum Landau gauge in the naive continuum limit.

When computing expectation values of some observable $O$, gauge fixing is implemented by constraining the integrations to those values of the link variables $U_{i,\mu}$ which satisfy the gauge fixing equations \eqref{eq:gaugefixing_gen}. Technically this can be done by a procedure named after Faddeev and Popov \cite{Faddeev:1967fc}, resulting in the expression
\begin{equation}\label{eq:gaugefixedexpectationvalue}
\langle O\rangle=\frac{\int \prod_{i,\mu}d U_{i,\mu}\Delta(g)\delta (f(g)) \exp(-S[U])O[U]/Z_\text{GF}}{\int \prod_{i,\mu}d U_{i,\mu}\Delta(g)\delta (f(g))\exp(-S[U])/Z_\text{GF}}
\end{equation}
for the expectation value of a gauge invariant observable $O[U]$. The so-called Faddeev-Popov determinant $\Delta=\det \mathcal{H}$ in this formula is the determinant of the Hessian matrix $\mathcal{H}$ of $F_U$ with elements
\begin{equation}
\mathcal{H}_{ij}(g)=\frac{\partial^2 F_U(g)}{\partial g_i \partial g_j},
\end{equation}
evaluated at the stationary points $g^{\text{s}}$ of $F_U$. The gauge fixing partition function occurring in the expectation value \eqref{eq:gaugefixedexpectationvalue} can be written in the form
\begin{equation}\label{eq:Z_GF}
Z_{\text{GF}}=\sum\sgn\Delta(g^{\text{s}}),
\end{equation}
where the summation is over all stationary points of $F_U$ \cite{Schaden:1998hz}.

In the 1970s, Gribov noticed that gauge fixing can in general not remove the ambiguity in the field variables completely \cite{Gribov:1977wm}. Instead, even after fixing a local gauge, there exist gauge-equivalent field configurations. In our setting, these so-called Gribov copies manifest themselves in multiple stationary points of the gauge fixing functional \eqref{eq:general_l_g_functional}. Despite the Gribov ambiguity, the Faddeev-Popov procedure for deriving the gauge-fixed expectation value works fine in the continuum, but a serious problem arises for compact U(1) and SU($N_c$) gauge theories on the lattice: Neuberger observed in the 1980s that, in these cases, the Gribov ambiguity leads to a vanishing lattice gauge fixing partition function $Z_\text{GF}$ \cite{Neuberger:1986vv,Neuberger:1986xz}, and hence to vanishing denominators in the gauge-fixed expectation value \eqref{eq:gaugefixedexpectationvalue}. From this so-called Neuberger problem, one has to conclude that the standard lattice Landau gauge \eqref{eq:general_l_g_functional} is not a suitable gauge fixing condition for compact U(1) and SU($N_c$) gauge theories on the lattice. This failure can be viewed as originating from a deeper cause: for compact U(1) or SU($N_c$) gauge groups, the standard lattice Landau gauge fixing procedure cannot restore a Becchi-Rouet-Stora-Tyutin (BRST) symmetry on the lattice.

In fact, Neuberger first observed a vanishing $Z_\text{GF}$ when he tried to establish BRST symmetry on the lattice. Later, Schaden interpreted the problem in terms of Morse theory and showed in \cite{Schaden:1998hz} that, by virtue of the Poincar\'{e}-Hopf theorem, one can write the gauge fixing partition function as
\begin{equation}\label{eq:Z_GF_is_Euler_char}
 Z_\text{GF} = \chi(G^N) = \left[\chi(G)\right]^N,
\end{equation}
where $\chi(G)$ denotes the Euler characteristic of the group manifold $G$. The Euler characteristic is a topological invariant, and for both, U(1) and SU($N_c$) group manifolds, we have $\chi(G)=0$, and $Z_\text{GF}$ therefore vanishes. Making use of such a topological interpretation, it was argued in \cite{vonSmekal:2007ns} that the Neuberger problem for an SU($N_c$) lattice gauge theory actually lies in the compact U(1) subgroup of SU($N_c$). Therefore, finding a solution to the U(1) Neuberger problem is expected to have an even bigger impact, as it should allow to evade the problem also for SU($N_c$) gauge groups.

In the next section, we will use the stationary points computed in section \ref{sec:stationarypoints} to explicitly demonstrate the Neuberger problem for a compact U(1) gauge theory on a one-dimensional lattice.

\subsection{One-dimensional compact U(1) lattice gauge theory}
\label{sec:gaugefixing}

Topologically, the group manifold of a compact U(1) gauge group is a circle. We parametrize this circle by means of angular variables, the link angles $\phi_{i,\mu}\in (-\pi,\pi]$ and the gauge transformation angles $\theta_{i}\in (-\pi,\pi]$. In terms of these angles, we can write the link variables as $U_{i,\mu} = \exp(\ii \phi_{i,\mu})$ and the gauge transformations as $g_{i}=\exp(\ii \theta_{i})$. Inserting these expressions into equation \eqref{eq:general_l_g_functional} and specializing to a one-dimensional lattice, the lattice Landau gauge fixing functional is identical to the function $F_\phi$ defined in \eqref{eq:F_phi}. This allows us to use the stationary points and Hessian determinants obtained in section \ref{sec:stationarypoints} for computing the gauge fixing partition function \eqref{eq:Z_GF} of a one-dimensional compact U(1) lattice gauge theory.

For even $N$, the computation of $Z_{\text{GF}}$ is particularly simple and does not even require explicit solutions for the stationary points: For every stationary point $s=(s_1,\dots,s_N)$ satisfying the gauge fixing equations \eqref{eq:s_gaugefixing} and \eqref{eq:s_gaugefixing_boundary}, there is a gauge transformed Gribov copy $s'= (s_1+\pi,\dots,s_N+\pi)$, likewise satisfying the gauge fixing equations. From equation \eqref{eq:Hessian_determinant} we can infer that the Faddeev-Popov determinants (or Hessian determinants) at $s$ and $s'$ have opposite signs,
\begin{equation}\label{eq:symm}
 \det \mathcal{H}(s') =\sum_{j=1}^{N} \prod_{\substack{k=1\\k\neq j}}^{N}\cos (s_k+\pi) =-\sum_{j=1}^{N} \prod_{\substack{k=1\\k\neq j}}^{N}\cos s_k = -\det \mathcal{H}(s).
\end{equation}
This symmetry allows us to easily read off that the gauge fixing partition function
\begin{equation}
Z_{\text{GF}}=\sum\sgn\Delta(\theta^\text{s})=0
\end{equation}
vanishes for the one-dimensional compact U(1) lattice gauge theory with an even number $N$ of lattice sites.

For odd $N$, no symmetry of the type \eqref{eq:symm} exist. In order to evaluate the gauge fixing partition function \eqref{eq:Z_GF} in this case, we make use of the Hessian determinant evaluated at the stationary points \eqref{eq:HessianDetAtSP}, as computed in section \ref{sec:stationarypoints}. Since $(\cos s_N)^{(N-1)}$ is positive for odd $N$, the sign of \eqref{eq:HessianDetAtSP} is determined by the remaining terms
\begin{equation}\label{eq:remainingterms}
\Biggl(1+\sum_{k=1}^{N-1}(-1)^{q_k}\Biggr)\Biggl(\prod_{j=1}^{N-1}(-1)^{q_j}\Biggr).
\end{equation}
This expression evidently does not depend on the parameter $l$ labelling a particular stationary point, nor on the specific sequence of binary parameters $q_k\in\{0,1\}$, but only on the number
\begin{equation}
n=\sum_{k=1}^{N-1} q_k
\end{equation}
of ``1''s in such a sequence. Writing \eqref{eq:remainingterms} in terms of $n$, we obtain $(N-2n)(-1)^n$, and the sign of this expression determines the sign of the Hessian determinant at a given stationary point. Since there are $\binom{N-1}{n}$ sequences $(q_1,\dots,q_{N-1})$ with a given $n$, times $|N-2n|$ different parameters $l$, we obtain
\begin{equation}
Z_{\text{GF}} = \sum\sgn\Delta(\theta^\text{s}) = \sum_{n=0}^{N-1}\binom{N-1}{n}|N-2n|\sgn(N-2 n)(-1)^{n} = \sum_{n=0}^{N-1}\binom{N-1}{n}(N-2n)(-1)^{n}=0.
\end{equation}
This result illustrates the occurrence of the Neuberger problem, i.e.\ of a vanishing gauge fixing partition function, for a one-dimensional compact U(1) lattice gauge theory. Since this leads to vanishing denominators in expectation values, and therefore to meaningless expressions, we must conclude that the chosen gauge fixing procedure is not adequate for this gauge theory.

\subsection{Modified lattice Landau gauges}
\label{sec:modifiedLLG}
In order to avoid the Neuberger problem, we might want to modify the gauge fixing condition \eqref{eq:gaugefixing} in some way. Apart from avoiding the Neuberger problem, there are further properties a modified gauge fixing condition should satisfy. In particular, the computational complexity should be such that a numerical implementation is feasible in practice, and the gauge fixing partition function $Z_\text{GF}$ should be independent of the orbit $\phi$. A few possible modifications of the lattice Landau gauge are discussed in this section.

\paragraph{Minimal lattice Landau gauge}
We had seen in section \ref{sec:gaugefixing} that $Z_\text{GF}$ vanishes due to cancellations of Hessian determinants with opposite signs. An obvious strategy to avoid such cancellations consists in strengthening the gauge fixing constraint, for example by requiring not only a stationary point of the gauge fixing functional \eqref{eq:F_phi}, but a local minimum \cite{Gribov:1977wm}. Apart from avoiding the Neu\-ber\-ger problem, this so-called minimal lattice Landau gauge is also numerically advantageous, as minima are easier to track down than general stationary points. However, as observed in theorem \ref{theorem} and the subsequent discussion, the number of minima depends on the gauge orbit $\phi$. Hence, also the gauge fixing partition function $Z_\text{GF}$ is orbit-dependent, and this is an undesired effect.

\paragraph{Absolute lattice Landau gauge}
The unwanted dependence of $Z_\text{GF}$ on the gauge orbit $\phi$ can generically be eliminated by further strengthening the gauge fixing constraint to the set of global minima of the gauge fixing functional \eqref{eq:F_phi}, and this choice is called absolute lattice Landau gauge. From \eqref{eq:F_at_SP} and theorem \ref{theorem} we find that the absolute minima of $F_\phi$ are given by the stationary points with $q_k=0$ and a parameter $l$ chosen such that $\cos[(\Phi+2\pi l)/N]$ is maximized.

Although the absolute lattice Landau gauge resolves both the Neuberger problem and, for generic values of $\Phi$, the problem of orbit-dependent $Z_\text{GF}$, it suffers from another drawback: In higher dimensions, where no analytical results are available, finding the global minimum of the gauge fixing functional is a computationally very hard problem, rendering an application of this gauge fixing condition practically impossible.

\paragraph{Stereographically modified lattice Landau gauge}
Another attempt to overcome the Neuberger problem, based on a stereographic projection of the U(1) group manifold, has been proposed and studied recently \cite{vonSmekal:2007ns,vonSmekal:2008es,Mehta:2009}. 
The stereographically modified lattice Landau gauge fixing functional for the compact U(1) case in one dimension can be written as
\begin{equation}\label{eq:Ftilde}
\tilde{F}_{\phi}(\theta) = -2 \sum_{k=1}^{N}\ln \left(\frac{1+\cos (\phi_{k} + \theta_{k+1} - \theta_{k})}{2}\right).
\end{equation}
For this function one can show that the number of stationary points is independent of the orbit $\phi$ and exponentially (in $N$) smaller compared to the standard lattice Landau gauge \cite{vonSmekalpvt}. Moreover, after removing the global gauge freedom, the Hessian determinant is generically positive definite, cancellation of signs of determinants cannot occur, and the Neuberger problem is therefore absent \cite{Mehta:2009,Mehta:to_appear_MLLG}. This latter property also generalizes to gauge groups SU($N_{c}$) and to higher dimensional lattices. It remains, however, an open question whether the desired property of orbit-independence of $Z_\text{GF}$ generalizes to higher-dimensional versions of the functional \eqref{eq:Ftilde}. In summary, the stereographically modified lattice Landau gauge satisfactorily solves the Neuberger problem on a one-dimensional lattice, but it remains to be seen whether this is true also in higher dimensions.


\section{Random-phase $XY$ model}
\label{sec:XY}

In this section, some recent results are reviewed that relate stationary points of the Hamiltonian function of a classical many-body system to the occurrence of a phase transition in the thermodynamic limit of large system size. The function $F_\phi$ defined in \eqref{eq:F_phi} is identified as the Hamiltonian of the random phase $XY$ model in one spatial dimension. This connection allows us to use the stationary points of $F_\phi$ and their Hessian determinants computed in section \ref{sec:stationarypoints} for an analysis of certain thermodynamic properties of the model.

\subsection{Stationary points and phase transitions}
\label{sec:phasetransitions}

As is long known, the stationary points of a classical Hamiltonian function can be employed to calculate or estimate certain physical quantities of interest. Famous examples include transition state theory \cite{Eyring:35} or Kramers's reaction rate theory for the thermally activated escape from metastable states \cite{Kramers40}, where the barrier height, i.\,e.\ the potential at a certain stationary point of the potential energy function, plays an essential role. More recently, the noise-free escape from quasi-stationary metastable states, whose lifetimes diverge with the system size, has been related to the presence of stationary points of marginal stability \cite{Tamarit_etal05}. Apart from studies of dynamical properties, stationary points have also been extensively used for estimating thermodynamical properties by means of the superposition approach \cite{StrodelWales08}.

Dynamical properties like the aforementioned ones are, as one might expect, not unrelated to the statistical physical behaviour of a system. Accordingly, as worked out beautifully in \cite{CaPeCo:00}, properties of stationary points\footnote{The authors of \cite{CaPeCo:00} discuss topology changes of constant-energy manifolds in phase space. Via Morse theory, such topology changes can be related to the stationary points of a sufficiently smooth Hamiltonian.} reflect in dynamical and statistical physical quantities simultaneously. This observation sparked quite some research activity, reviewed in \cite{Kastner:08}, with the aim of relating equilibrium phase transitions to stationary points and their indices. Among other things, it was observed that stationary points of the potential energy function generate nonanalyticities of the microcanonical entropy like the ones reported in \cite{CaKa:06}. Subsequently, it was noticed in \cite{KaSchneSchrei:07,KaSchne:08,KaSchneSchrei:08} that the Hessian determinant of the Hamiltonian, evaluated at the stationary points, adds a crucial piece of information that helps to discriminate whether or not a phase transition occurs. Omitting the technical details, the essence of the criterion on the Hessian determinant can be captured as follows \cite{NardiniCasetti09}.
\begin{hdc}
Let
\begin{equation}\label{eq:Hstandard}
H(p,q)=\frac{1}{2}\sum_{k=1}^N p_k^2 + V(q_1,\dots,q_N)
\end{equation}
be the Hamiltonian of a system with $N$ degrees of freedom, possessing a number of stationary points that grows at most exponentially with $N$. Here, $p=(p_1,\dots,p_N)$ and $q=(q_1,\dots,q_N)$ denote, respectively, the vectors of momenta and positions, and $V$ is the potential energy. In the thermodynamic limit $N\to\infty$, a phase transition can occur at some critical energy per degree of freedom $e_\text{c}$ only if the following two conditions are met:
\begin{enumerate}
\item There exists a sequence $\bigl\{q_\text{s}^N\bigr\}_{N=N_0}^\infty$ of stationary points of $V$ such that
\begin{equation}
v_\text{c}:=\lim_{N\to\infty}\frac{V\bigl(q_\text{s}^N\bigr)}{N}
\end{equation}
converges and equals $\langle v\rangle(e_\text{c})$ (i.e.\ the ensemble expectation value of $v=V/N$ at the energy $e_\text{c}$).
\item The asymptotic behaviour of the Hessian matrix $\mathcal{H}$ of V, evaluated at the critical points $q_\text{s}^N$ contained in that sequence, is such that
\begin{equation}
\lim_{N\to\infty}\bigl|\det \mathcal{H}\bigl(q_\text{s}^N\bigr)\bigr|^{1/N}=0.
\end{equation}
\end{enumerate}
\end{hdc}
Note that this criterion is necessary, but not sufficient, for a phase transition to occur: finding a sequence of stationary points with the behaviour specified above does not guarantee a transition to take place at the corresponding critical energy. However, as model calculations show, the criterion usually appears to single out precisely the correct transition energies \cite{KaSchne:08,KaSchneSchrei:08}. Importantly for the application of the above criterion, knowledge of a suitably chosen subset of all the stationary points of $V$ may be sufficient. This matter of fact was pointed out by Nardini and Casetti \cite{NardiniCasetti09}, and suitably constructed sequences of stationary points were used to single out the phase transition and determine its critical energy for a model of gravitating masses.

Comparing the above criterion to other analytic tools in the statistical physics of phase transitions, a remarkable property is its {\em locality}\/ in configuration space. In contrast to, say, the calculation of a partition function, no averaging over a large, high-dimensional manifold is necessary. Instead, only the local properties of a sequence of stationary points needs to be analyzed. Of course, finding an appropriate sequence of stationary points can be equally hard or impossible, but in certain instances such a local approach may prove beneficial.

\subsection{Random phase $XY$ model}
\label{sec:PTXY}

The study of spin models has been very fruitful towards the understanding of phase transitions. First, such models are typically simpler than continuum systems, and second they are successfully employed for the modelling of condensed matter phenomena like ferromagnetism and others. An exceptionally popular class of such models are $O(n)$ vector models, consisting of $n$-dimensional unit vectors $\sigma_k$ placed on the sites of a lattice. The interaction between the classical spin vectors is described by a Hamiltonian of the form
\begin{equation}
H(\sigma)=J\sum_{\langle k,l\rangle}(1-\sigma_k\cdot\sigma_l)
\end{equation}
with coupling constant $J$. The angular brackets $\langle \cdot,\cdot\rangle$ denote a summation over all pairs of nearest-neighbouring sites on the lattice. $O(n)$ vector models include as special cases (a) the Ising model, $n=1$, (b) the $XY$ model, $n=2$, and (c) the Heisenberg model, $n=3$. For $n=2$, we can parametrize the two-dimensional unit vectors $\sigma_k$ by an angular variable $\theta_k$, obtaining the Hamiltonian of the classical $XY$ model,
\begin{equation}\label{eq:H_XY}
H(\theta)=J\sum_{\langle k,l\rangle}[1-\cos(\theta_k-\theta_l)].
\end{equation}

A number of disordered variants of $O(n)$ vector models can be found in the literature, and their study has improved the understanding of glassy behaviour, ageing, and related phenomena. One such disordered variant, the random phase $XY$ model, is characterized by the Hamiltonian
\begin{equation}
H_\phi(\theta)=J\sum_{\langle k,l\rangle}[1-\cos(\phi_l+\theta_k-\theta_l)],
\end{equation}
where the $\phi_l$ are angular variables, drawn randomly according to some probability distribution. On a one-dimensional lattice and with coupling $J=1$, this Hamiltonian takes on the form of $F_\phi$ as defined in \eqref{eq:F_phi}.
 
\begin{figure}\center
\includegraphics[width=50mm]{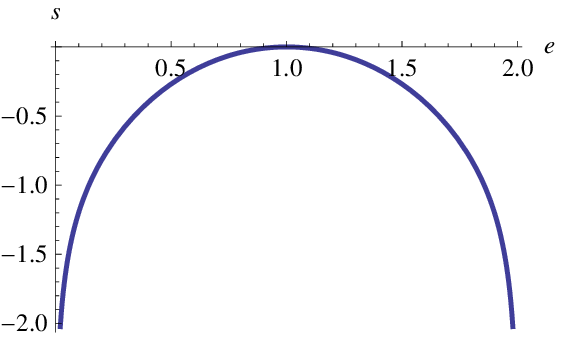}
\hspace{15mm}
\includegraphics[width=50mm]{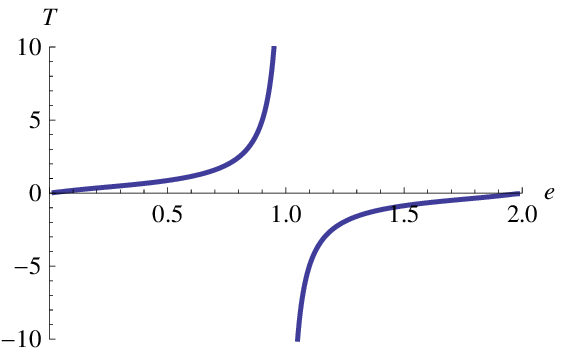}
\vspace{-10mm}
\caption{\label{fig:s_of_e}
Graph of the entropy $s$ and the temperature $T$, both as functions of the energy $e$, for the one-dimensional $XY$ model. The range of accessible energies is $[0,2]$, where energies $0<e<1$ correspond to positive temperatures, energies $1<e<2$ to negative temperatures.
}
\end{figure}

On a lattice of spatial dimension $d=2$ or higher, the random phases $\phi_k$ cause frustration, resulting in a number of interesting physical properties. In one dimension, however, frustration does not occur, and the random phases do not make much of a difference compared to the ordered model \eqref{eq:H_XY}. We will therefore mainly focus on the one-dimensional $XY$ model in the absence of disorder. The Hamiltonian of this model is
\begin{equation}\label{eq:H_XY_1d}
H(\theta)\equiv F_0(\theta)=\sum_{k=1}^N [1-\cos(\theta_{k+1}-\theta_k)],
\end{equation}
i.\,e.\ $F_\phi$ as defined in \eqref{eq:F_phi} with $\phi_k=0$ for all $k=1,\dots,N$. An exact solution for the canonical partition function (or, equivalently, for the canonical free energy $f$) as a function of the inverse temperature $\beta$ has been reported in \cite{Stanley69}. In the thermodynamic limit $N\to\infty$ and translated into our notation, the solution is given by
\begin{equation}
f(\beta)=1-\frac{1}{\beta}\ln I_0(\beta),
\end{equation}
where $I_0$ denotes the modified Bessel function of the first kind. For our purposes, the entropy $s$ as a function of the energy $e$ is more convenient, and can be obtained from $f$ by means of a Legendre transform,
\begin{equation}
s(e)=\bar{\beta}(e-1)+\ln I_0(\bar{\beta}(e)),
\end{equation}
where $\bar{\beta}(e)$ is the solution of
\begin{equation}
I_1(\bar{\beta})=(1-e)I_0(\bar{\beta}).
\end{equation}
The energy $e$ can take on values from the interval $[0,2]$. The graph of $s$ is shown in figure \ref{fig:s_of_e}, together with a plot of the temperature $T(e)=[s'(e)]^{-1}$.

Although this model has no phase transition, it will prove instructive as what regards the study of stationary points of many-body systems in the spirit of section \ref{sec:phasetransitions}.

\subsection{Stationary points analysis of the one-dimensional $XY$ model}
\label{sec:statpoint1dXY}

To study the one-dimensional $XY$ model in the spirit of the criterion stated in section \ref{sec:phasetransitions}, we need to compute, for suitable sequences of stationary points $\theta^\text{s}$, the (potential) energy $e_N(\theta^\text{s})=F_0(\theta^\text{s})/N$ together with the scaled Hessian determinant,
\begin{equation}
\mathcal{D}_N(\theta^\text{s}):=\left|\det\mathcal{H}(\theta^\text{s})\right|^{1/(N-1)}.
\end{equation}
Inserting the stationary points specified by \eqref{eq:s_solutions2} with $s_N$ given by \eqref{eq:s_final} into the Hamiltonian \eqref{eq:H_XY_1d}, we obtain
\begin{equation}
e_N(\theta^\text{s})=1-\frac{1}{N}\biggl(1+\sum_{k=1}^{N-1}(-1)^{q_k}\biggr)\cos s_N.
\end{equation}
For the scaled Hessian determinant, we find 
\begin{equation}
\mathcal{D}_N(\theta^\text{s})=|\cos s_N|\,\Bigl|\sum_{k=1}^{N-1}(-1)^{q_k}\Bigr|^{1/(N-1)}
\end{equation}
by making use of the expression \eqref{eq:HessianDetAtSP} for the Hessian determinant.

\begin{figure}\center
\psfrag{D}{$\scriptstyle \mathcal{D}$}
\includegraphics[width=50mm]{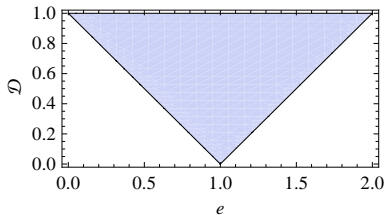}
\vspace{-2mm}
\caption{\label{fig:D-e-plane}
In the plane of energy per degree of freedom $e$ and scaled Hessian determinant $\mathcal{D}$, the region is shown for which stationary points $\theta^\text{s}$ with the corresponding values $(e,\mathcal{D})$ exist. In the thermodynamic limit $N\to\infty$, the stationary points fill the shaded region densely.
}
\end{figure}

To apply the criterion of section \ref{sec:phasetransitions}, we need to choose appropriate sequences $\{\theta^\text{s}\}_{N=N_0}^\infty$ of stationary points such that the corresponding sequence of energies $e_N(\theta^\text{s})$ converges. This is most conveniently done by fixing values of $s_N$ and of the parameter
\begin{equation}
[-1,1]\ni l_q:=\frac{1}{N}\biggl(1+\sum_{k=1}^{N-1}(-1)^{q_k}\biggr).
\end{equation}
For any sequence%
\footnote{To construct a sequence of stationary points with a fixed value of $s_N$, it is in general necessary to restrict the sequence to some infinite subset of system sizes $N$ for which solutions with the chosen value of $s_N$ exist.}
 of stationary points with fixed $s_N$ and $l_q$, we can write
\begin{equation}\label{eq:e_N}
e_N(\theta^\text{s})=1-l_q\cos s_N
\end{equation}
and
\begin{equation}
\mathcal{D}_N(\theta^\text{s})=|\cos s_N||Nl_q-1|^{1/(N-1)}.
\end{equation}
In the thermodynamic limit $N\to\infty$, the latter simplifies to
\begin{equation}\label{eq:D}
\mathcal{D}:=\lim_{N\to\infty}\mathcal{D}_N(\theta^\text{s})=\left|\cos s_N\right|,
\end{equation}
becoming independent of $l_k$. Note that, by definition, $s_N$ is fixed for the sequence of stationary points we are considering and is therefore not affected by the limiting procedure in \eqref{eq:D}. The possible values of $s_N$, as determined by \eqref{eq:s_final}, become dense on the interval $(-\pi,\pi]$ in the thermodynamic limit. Combining equations \eqref{eq:e_N} and \eqref{eq:D}, we obtain
\begin{equation}\label{eq:D_of_e}
\mathcal{D}=\left|\frac{1-e_N(\theta^\text{s})}{l_q}\right|.
\end{equation}
This implies that, for all stationary points $\theta^\text{s}$, the pairs $(e_N,\mathcal{D})$ are located within the shaded triangle plotted in figure \ref{fig:D-e-plane}, and in fact they fill the triangle densely in the limit $N\to\infty$.

Interpreting these results with regard to the criterion stated in section \ref{sec:phasetransitions}, we first have to note that, in contrast to  \eqref{eq:Hstandard}, the Hamiltonian of the one-dimensional $XY$ model has no kinetic energy term $\frac{1}{2}\sum_{k=1}^N p_k^2$. Luckily, this is of minor importance and the criterion still applies, with the only simplification that critical energies $e_\text{c}$ and critical potential energies $v_\text{c}$ coincide (see \cite{CaKaNe09} for a related discussion).

From the fact that $\mathcal{D}>0$ for all $e\neq1$ we can conclude that the one-dimensional $XY$ model has no phase transition for these values of the energy. Only for $e=1$, the criterion cannot exclude a phase transition. Of course, we know from the thermodynamic solution that no such transition exists. As pointed out in section \ref{sec:phasetransitions}, the criterion on the Hessian determinant allows to exclude a phase transition when $\mathcal{D}$ is bounded away from zero, but a vanishing $\mathcal{D}$ does not necessary imply the occurrence of a phase transition at the corresponding energy $e$.\footnote{A vanishing $\mathcal{D}$ in the absence of a phase transition had been observed earlier for a mean-field model of two particle species \cite{SantonCoutinhoFilho09}.} Moreover, note that $e=1$ is a very peculiar energy value of the one-dimensional $XY$ model: This energy corresponds to temperatures $T=\pm\infty$, as is evident from the right plot in figure \ref{fig:s_of_e}. So even if the entropy $s(e)$ had shown a nonanalyticity at $e=1$, the corresponding phase transition would have occurred at infinite temperature.

Although we could not expect much news on the well-known thermodynamics of the one-dimensional $XY$ model, there are a number of other interesting observations to be made: The plot in figure \ref{fig:D-e-plane} shows a remarkable difference compared to earlier work in which phase transitions had been analyzed via the Hessian determinant at stationary points: For the mean-field type models studied in \cite{KaSchne:08,KaSchneSchrei:08,SantonCoutinhoFilho09}, a unique value of $\mathcal{D}$ was found for each value of the energy $e$. In contrast, for the one-dimensional $XY$ model, we observe $\mathcal{D}$-values densely covering an interval. This property, which we assume to be generic, has remarkable consequences for the application of the Hessian determinant criterion: It is, apparently, not sufficient to study {\em any}\/ sequence $\{\theta^\text{s}\}_{N=N_0}^\infty$ for which the sequence of energies $e_N(\theta^\text{s})$ converges. Instead, in order to single out energies for which $\mathcal{D}$ can become zero, a suitable sequence of stationary points has to be employed. To get an idea what sequences are the relevant ones, the following considerations are of help.

In section \ref{sec:stationarypoints}, we had remarked that, among the many stationary points of the function $F_\phi$, there exist two classes of particularly simple ones: (a) The solutions $(s_1,\dots,s_N)$ with $s_k\in\{0,\pi\}$ for all $k$. These solutions, by virtue of equation \eqref{eq:D}, all yield $\mathcal{D}=1$, and they densely populate the upper boundary of the triangle in figure \ref{fig:D-e-plane}. (b) A second class of simple solutions consists of identical values $s_k=2\pi l/N$ for all $k$, where $l$ is some integer between 1 and $N$. These solutions all correspond to $q_k=0$ for all $k=1,\dots,N-1$, and therefore $l_q=1$. By means of equation \eqref{eq:D_of_e}, the solutions are found to be located on the two equal edges of the isosceles triangle in figure \ref{fig:D-e-plane}, and they provide lower bounds on $\mathcal{D}$ for all accessible energies $e\in[0,2]$.

Hence, for the purpose of singling out those energies for which the rescaled Hessian determinant $\mathcal{D}$ can be vanishing, this second class of solutions with $s_k=2\pi l/N$ for all $k$ would have been sufficient. What is more, these symmetric solutions are particularly easy to analyze. Maybe this points towards a deeper reason behind the remarkable success of Nardini and Casetti's analysis of a model of gravitating masses on a ring, using only a subset of symmetric stationary points \cite{NardiniCasetti09}.

Another interesting observation arises when comparing results for the stationary points of $F_\phi$ obtained for different boundary conditions. For anti-periodic boundary conditions, $s_{N+1} = -s_{1}$ and $s_{0} = -s_{N}$, a calculation of the stationary points has been reported in \cite{vonSmekal:2007ns,Mehta:2009}. In this case, all $2^N$ stationary points are of type (a), i.e.\ $(s_1,\dots,s_N)$ with $s_k\in\{0,\pi\}$. Plotting $\mathcal{D}$ as a function of the energy $e$, only the horizontal line $\mathcal{D}=1$ in figure \ref{fig:D-e-plane} is obtained. According to the criterion of section \ref{sec:phasetransitions}, this result rules out the existence of a phase transition in the thermodynamic limit. Since thermodynamic quantities of short-range interacting systems are known to be independent of the boundary conditions, we can infer the absence of a phase transition for the one-dimensional nearest-neighbour $XY$ model with any boundary conditions from the stationary points in the case of anti-periodic boundary conditions. Again, this observation might prove useful when studying other models: Since the boundary conditions strongly affect the stationary points, a suitable choice of the boundary conditions might simplify the computation in some cases, and/or allow to establish the criterion of  section \ref{sec:phasetransitions} in others.

Let us finally remark that stationary points of the one-dimensional $XY$ model and their relation to phase transitions had been analyzed previously by Casetti {\em et al.}\ in section 3 of \cite{CaPeCo:03}. There are, however, a number of important differences to our work. First, these authors employ a different strategy to deal with the global invariance $\theta_k\to\theta_k+\alpha$. Instead of fixing one variable $\theta_N=0$ as in our analysis, they introduce a symmetry-breaking external magnetic field in the Hamiltonian. Second, the study in \cite{CaPeCo:03} focusses on stationary points and their indices, but does not consider the Hessian determinant at these points. This is not much of a surprise since, at that time, the importance of the Hessian determinant at the stationary points had not yet been realized. Third, Casetti {\em et al.}\ consider those special stationary points where $s_k\in\{0,\pi\}$ to be exhaustive, missing out the many other ones. Considering this subset of stationary points only, the vanishing rescaled Hessian determinant $\mathcal{D}$ at $e=1$ would have gone unnoticed.  


\section{Summary and Outlook}
\label{sec:summary}
We have studied the function $F_\phi (\theta)=\sum_{k=1}^N [1-\cos(\phi_k +\theta_{k+1}-\theta_k)]$ with periodic boundary conditions, obtaining all its stationary points for any odd $N$. We have employed this result to illustrate the relevance and usefulness of stationary points in different branches of physics.

When interpreting $F_\phi$ as the lattice Landau gauge fixing functional in one-dimensional compact U(1) lattice gauge theory, we used the exact solution for the stationary points $\theta^\text{s}$ (called Gribov copies in this context) and their Hessian determinants (or Faddeev-Popov determinants) $\Delta(\theta^\text{s})$ to show that the gauge fixing partition $Z_\text{GF}=\sum\sgn\Delta(\theta^\text{s})$ vanishes. The Neuberger problem of vanishing $Z_{GF}$ results in ill-defined expectation values, pointing towards a deeper problem: the standard lattice Landau gauge fixing fails to establish the BRST symmetry on the lattice, and is therefore not adequate for the one-dimensional compact U(1) lattice gauge theory.
Also based on the exact expression for the stationary points, a number of strategies of how to avoid the Neuberger problem and the Gribov ambiguity were discussed. The hope is that a thorough understanding of the one-dimensional case is helpful for devising strategies of how to avoid these problems also for gauge theories on higher dimensional lattices.

Interpreting $F_\phi$ as the Hamiltonian of the one-dimensional random phase $XY$ model in classical statistical physics, the solutions for the stationary points and their Hessian determinants allowed us to evaluate a criterion which makes predictions on the existence of phase transitions and their critical energies in the thermodynamic limit. In particular, unlike the long-range interacting models analyzed earlier, the computation provides the first example of a model where, for a given value of the energy $e$, the thermodynamic limit $\mathcal{D}$ of the Hessian determinant densely covers a range of values. 
Furthermore, we observed that the boundary conditions have a drastic effect on the stationary points, and therefore on the values of $\mathcal{D}$. These findings, we believe, can be helpful when studying more complicated models where a complete solution for the stationary points is not feasible. 

For lattices of dimension two or larger, the $XY$ model is known to undergo a phase transition. The $XY$ model on a two-dimensional square lattice therefore appears to be an ideal choice for generalizing the methods and results of the present article. In particular, solutions of the form \eqref{eq:s_solutions} straightforwardly generalize to higher-dimensional lattices. However, for dimensions larger than one they cannot be expected to be the only solutions of the stationary point equations, but many others might exist. Apart from such an analytical approach, one of us is currently working on numerical approaches for computing the stationary points of $XY$ models in dimensions larger than one \cite{Mehta:2009zv,Hughes:to_appear}. In order to check whether all stationary points have been found in such a computation, one can use the Neuberger zero as a necessary, though not sufficient, condition.

Apart from the interpretation of $F_\phi$ as a Hamiltonian function, its stationary points can also be used to discuss non-Hamiltonian dynamical systems: The stationary points of $F_0$ can be shown to be the fixed points of the one-dimensional Kuramoto model with nearest-neighbour interactions,
\begin{equation}
\frac{\dd \tilde{\theta}_k}{\dd t} = \omega_k + \frac{K}{2}\left[\sin(\tilde{\theta}_{k+1}-\tilde{\theta}_k) - \sin(\tilde{\theta}_k-\tilde{\theta}_{k-1})\right],
\end{equation}
provided the frequencies $\omega_k=\omega$ are identical for all $k=1,\dots,N$. Under the latter condition, one can transform the equations of motion to new variables $\theta_k=\tilde{\theta}_k-\omega t$, yielding
\begin{equation}\label{eq:1dKuramoto}
\frac{\dd \theta_k}{\dd t} = \frac{K}{2}\left[\sin(\theta_{k+1}-\theta_k) - \sin(\theta_k-\theta_{k-1})\right],
\end{equation}
where $K\in\RR$ is some coupling constant. At a stationary point $\theta^\text{s}$ of $F_0$, the right-hand side of \eqref{eq:1dKuramoto} vanishes, resulting in a fixed point of the dynamics. The fixed point is stable if $\theta^\text{s}$ is a minimum of $F_0$ as characterized in Theorem \ref{theorem}. Although the one-dimensional nearest-neighbour Kuramoto model is known to show no synchronization transition \cite{0305-4470-21-13-005}, its fixed points may be useful for studying synchronization-related dynamical features of finite systems \cite{OchabGora}.

We conclude this outlook by mentioning another example of a lattice gauge fixing functional that can be interpreted as a physical system: It was observed in \cite{Cucchieri:2009kk} that the lattice gauge fixing functional of the linear covariant gauge can be interpreted as the Hamiltonian of a spin-glass model in a random external magnetic field. For a compact U(1) gauge group, the stationary points of this gauge fixing functional also correspond to the fixed points of a Kuramoto model with nearest neighbour interactions and frequencies $\omega_k$ distributed according to some probability distribution. For other compact Lie groups, similar relations are expected to exist between the lattice gauge fixing functionals of the linear covariant gauge and non-Abelian generalizations of the Kuramoto model \cite{1751-8121-42-39-395101}. Similar to the results reported in the present article, a stationary point analysis may lead to interesting cross-connections linking the lattice gauge and dynamical system interpretations of these models.




\section*{Acknowledgments}
D.\,M.\ was supported by Science Foundation Ireland research grant number 08/RFP/PHY 1462, and most of this work has been done while D.\,M.\ was affiliated with the Department of Mathematical Physics, National University of Ireland Maynooth. 
M.\,K.\ acknowledges financial support by the {\em Incentive Funding for Rated Researchers}\/ programme of the National Research Foundation of South Africa. 

\bibliographystyle{model1a-num-names}
\bibliography{1dXY.bib}

\end{document}